\documentclass[12pt]{article}
\newcommand{\myskip}{\vspace{\baselineskip}}
\newcommand{\mysection}[1]{\par\myskip\noindent\textbf{#1}\myskip\par}

\begin{document}
\begin{center}
{\large\textbf{Reduced Symmetry of Heterointerfaces and Orientational Pinning of Quantum Hall Stripe Phase}} 

\myskip
E.E.~Takhtamirov$^\dag$ and V.A.~Volkov\\
Institute of Radioengineering and Electronics of RAS\\
Mokovaya 11, 101999, Moscow, Russia\\
$^\dag$ E-mail: takhtam@cplire.ru

\end{center}

\mysection{Abstract}

\myskip 
\parbox{4.5in}
{In a 2D electron system in (001) GaAs/AlGaAs upon filling high Landau levels, it was
recently observed a new class of collective states, which are recognized now to relate to the
spontaneous formation of a charge density wave (``stripe phase''). Here we analyze one of the
possible mechanisms of the stripe pinning along the crystallographic direction
[110]---the native effective mass anisotropy of 2D electrons. It is shown that
for a symmetric quantum well this anisotropy is a linear function of applied
per\-pen\-di\-cu\-lar-to-plane electric field. So, if it is the anisotropy that defines the
direction of the stripes, the critical in-plane magnetic field, when the stripes rotate,
will strongly depend on the bias. The proper experiment will answer the question whether the
native anisotropy of the effective mass is the mechanism of the pinning.}

\mysection{Introduction}
The experiments by Lilly, et al. and Du, et al. [1] revealed a new class of collective
states of 2D electron system in excited Landau levels in the single heterojunction (001)
GaAs/AlGaAs. Now it is generally accepted that these states are related to spontaneous formation
of the unidirectional charge density wave (quantum Hall stripe phase) predicted and calculated
earlier [2].
This point of view was primarily based on the observation of a giant resistance anisotropy
in this system. The ratio of resistances along the crystallographic directions $[1\bar 10]$ and
$[110]$ reaches the values of $R_{xx}/R_{yy}\sim5$-$3500$, depending on the sample geometry,
where $[110]$ is the ``easy'' conductivity direction. Moreover, the behavior of all conductivity
tensor components qualitatively agrees with the theory [3].

It was shown that the in-plane magnetic field $B_{\|}\sim 1$\,T can
change the direction of easy conductivity [4]. It was made a conclusion that, at high
enough $B_{\|}$, the direction of easy conductivity is perpendicular to the ${\bf B}_{\|}$
direction. The theoretical analysis [5] of the influence of ${\bf B}_{\|}$ carried out in the
Hartree-Fock approximation partially explained the results. But among researchers still there is no
agreement on the native mechanism that makes the stripes be pinned coherently across the sample
along the crystallographic direction $[110]$ at $B_{\|}=0$.

It was supposed by Kroemer [6] that the reduced symmetry $C_{2v}$ of the heterojunction composed
of III-V semiconductors might be the origin of the orientational pinning of the quantum Hall
stripe phase. Adopting this idea we have recently showed [7] that for a typical single-interface
structure the effective masses along $[110]$ and $[1\bar 10]$ directions differ by about
$0.1\%$ (``native'' anisotropy). The in-plane magnetic field ${\bf B}_{\|}$ leads to the
``magnetic'' anisotropy of the effective mass ($\propto B^2_{\|}$). The 2D electron system with
the anisotropic mass and isotropic Coulomb interaction is equivalent to that one with the
isotropic (cyclotron) mass and anisotropic Coulomb interaction. We supposed that it was the
effective anisotropic interaction that pinned the orientation of the stripes and allowed to
detect the phase in the magnetotransport experiments. The interplay of the native and magnetic
contributions to the effective mass anisotropy leads at $B_{\|}\sim 0.5$\,T to the isotropic
electron spectrum if ${\bf B}_{\|}$ along $[110]$. As a result, the direction of the stripes
becomes chaotic and the conductivity turns isotropic. Further increase of $B_{\|}$ makes
magnetic anisotropy dominate and the stripes align along $[1\bar 10]$ (perpendicular to
${\bf B}_{\|}$ for the one-subband case, and parallel to ${\bf B}_{\|}$ for the two-subband case).
The results seem to be in quantitative agreement with the experimental findings.

Here we will consider the orientational pinning of the stripe phase in a square quantum well.
The goal of the paper is to find a procedure that will unambiguously give the answer on the
question whether the native anisotropy of the effective mass is the mechanism of pinning of
the quantum Hall stripe phase.

\mysection{Symmetric quantum well in perpendicular electric field}
For a symmetric quantum well in the perpendicular-to-plane magnetic field the crystallographic
directions $[110]$ and $[1\bar 10]$ are mutually equivalent.
The system has $D_{2d}$ symmetry, and the conductivity should be anisotropic only as an in-plane
magnetic field is applied. This is in agreement with the recent data by Pan, et al. [8].

We analyse the single-particle Hamiltonian for electron in a square quantum well with
heterointerfaces at $z=0$ and $z=L$. Let us assign $x \| [1 \bar 1 0]$, $y \| [110]$ and
$z \| [001]$, and introduce magnetic field ${\bf B}$ in the vector-potential gauge
${\bf A} = (B_y z, -B_x z + B_z x, 0)$. Following Refs.~[7,9] we may obtain the orbital
part of the 3D Hamiltonian:
\begin{eqnarray}
H_{3D}&=&\frac {p^2_z}{2m^*} + V(z) + \frac 12 \left( \frac 1 {m^*} -
\alpha_1 \delta \left( z\right) + \alpha_2 \delta \left( z-L\right) \right)
\left( p_x + \frac e c B_y z \right)^2\nonumber\\
& &+\frac 12 \left( \frac 1 {m^*} + \alpha_1
\delta \left( z\right) - \alpha_2 \delta \left( z-L\right) \right)
\left( p_y - \frac e c B_x z + \frac e c B_z x \right)^2.\label{h}
\end{eqnarray}
Here, ${\bf p}$ is the momentum operator, $m^*$ is the band edge effective mass, $V(z)$ is
the effective potential of the conduction band edge, $e$ is the elementary
charge, $c$ is the speed of light, $\delta (z)$ is the Dirac $\delta$-function. If the quantum
well is {\em true} symmetric, symmetric is the potential $V(z)$, and the interface parameters
$\alpha_1$ and $\alpha_2$, see Ref.~[7], are equal, $\alpha_1=\alpha_2=\alpha$.
Generally these parameters may differ as different are the growth conditions for the direct,
AlGaAs/GaAs, and inverted, GaAs/AlGaAs, heterointerfaces. 

For a 2D layer of finite thickness the in-plane component of magnetic field
can be treated perturbatively. To the second order in $B_{\|}$
this procedure brings about a diamagnetic shift of 2D subband energies, shift
of the center of the Landau orbit in 2D quasi-momentum space, and an increase (for the lowest
subband) in the effective mass in the direction perpendicular to ${\bf B}_{\|}$. The terms
proportional to $\alpha_i$ ($i=1,\,2$) in Eq.~(\ref{h}) also can be treated perturbatively. For
simplicity, we assume that ${\bf B}_{\|}$ is parallel to either $[1 \bar 1 0]$ or $[110]$,
so that $B_xB_y = 0$. Collecting all terms second-order in $B_{\|}$ and first-order in $\alpha_i$,
and one obtains the following expression for 2D Hamiltonian of the ground subband [7]:
\begin{eqnarray}
H ^1_{2D} = E_1 + \frac {e^2}{2m^*c^2} \left( B^2_x + B^2_y \right)
\left(\left\langle 1\mid z^2 \mid 1\right\rangle -
\left\langle 1\mid z \mid 1\right\rangle^2\right)\nonumber\\
 +\frac 1{2m^*} \left( 1 - \frac {\Delta _{\rm nat}}2 -
\frac {B_y^2}{B_{\|}^2} \Delta _{\rm B} \right)
\left( p_x + \frac e c B_y \left\langle 1\mid z \mid 1\right\rangle \right)^2\nonumber\\
 +\frac 1{2m^*} \left( 1 + \frac {\Delta _{\rm nat}}2 -
\frac {B_x^2}{B_{\|}^2} \Delta _{\rm B} \right)
\left( p_y +\frac e c B_z x -\frac e c B_x \left\langle 1\mid z \mid 1\right\rangle \right)^2 .
\label{h0}
\end{eqnarray}
Here $\mid m \rangle=\Phi_m (z)$ is $z$-motion envelope function of $m$th subband, it is taken
to be real. The parameters of the natural anisotropy of effective mass and its anisotropy
induced by the magnetic field are
\begin{eqnarray}
\Delta_{\rm nat} = 2 m^*\left\langle 1\mid \alpha_1 \delta (z) - \alpha_2 \delta (z-L)\mid 1
\right\rangle, \quad
\Delta_{\rm B} = \frac {2e^2B^2_{\|}}{m^*c^2}{\sum_m}^{\prime }
\frac {\left\langle 1\mid z \mid m\right\rangle^2}{E_m - E_1},\label{delty}
\end{eqnarray}
where $E_m$ is the energy of the bottom of the $m$th subband at $B=0$.

For a symmetric quantum well $\Delta_{\rm nat}=0$. Nevertheless, when the
per\-pen\-di\-cu\-lar-to-plane electric field ${\bf F}$ is applied, the system gets $C_{2v}$
symmetry and should behave like a single-interface one. That means $\Delta_{\rm nat} \ne 0$ being a
linear function of $F$. It is the consequence of the linear Stark effect, which arise
as the Hamiltonian (\ref{h}) does not possess the symmetry $H_{3D}(z)=H_{3D}(L-z)$. Consider
the energy $eFz$ as a perturbation; $\Phi_m$ will be unperturbed wave functions. Then
\begin{eqnarray}
\Delta_{\rm nat} = 4 m^* \alpha e F {\sum_m}^{\prime }
\frac {\left\langle 1\mid z \mid m\right\rangle}{E_m - E_1}
\left( \Phi_1(0)\Phi_m(0)-\Phi_1(L)\Phi_m(L)\right).
\label{e-ind}
\end{eqnarray}
Generally, for any structure $\Delta_{\rm nat}=C+D\cdot F$, where $C$ and $D$ are
some parameters. For a symmetric quantum well $C=0$.

Let us analyze some recent experimental results by Cooper, et al. [10]. In particular, for a
formally symmetric quantum well (001) ${\rm  GaAs/Al_{0.24}Ga_{0.76}As}$ they found that the
resistance was anisotropic even without the in-plane magnetic fieled. This result is in
contrast with data by Pan, et al. [8]. Two possible explanations of the contradictory may
be given. The first one relates to the difference of the
interface parameter $\alpha$ for the direct and inverted heterointerfaces in the sample of
Ref.~[10], albeit the quantum well was intended to be grown symmetric. Another explanation
is based on existence of unintentional per\-pen\-di\-cu\-lar-to-plane electric field,
which pushes the electron wave function onto one of the heterointerfaces and induces the
effective mass anisotropy (\ref{e-ind}).

We present some results of the selfconsistent Shr\"odinger-Poisson computing and comparison
with the experiment, Ref.~[10].
For the heterojunction ${\rm  GaAs/Al_{0.3}Ga_{0.7}As}$ from comparison
with the experiments we obtained the value $\alpha = 1.1\cdot 10^{20}$~cm/g [7].
The linear interpolation for the symmetric quantum well GaAs with bariers
${\rm Al_{0.24}Ga_{0.76}As}$ gives $\alpha_1=0.88\cdot 10^{20}$~cm/g. Assume the heterointerfaces
in the sample of Ref.~[10] not equivalent, and no unintentional
per\-pen\-di\-cu\-lar-to-plane electric field presents. Then from the condition
$\Delta_{\rm nat}=\Delta _{\rm B}$ at critical magnetic field $B_{\|}=0.24$~T, when the resistance
becomes isotropic, we get $\alpha_2=-3.1\cdot 10^{20}$~cm/g. This value looks strange,
but no one can exclude such a case. Suppose now the quantum well microscopically symmetric,
$\alpha_1=\alpha_2=0.88\cdot 10^{20}$~cm/g. Then the electric field $F=1.5\cdot 10^4$~V/cm is
able to induce the effective mass anisotropy that will be cancelled in the magnetic field
$B_{\|}=0.24$~T. Note, that the sum in (\ref{delty}) rapidly diminishes with decreasing the
width of the quantum well (while $L=300$~\AA \, in Ref.~[10]).  If the native anisotropy of the
effective mass exists, the critical in-plane magnetic field will be higher for narrower
quantum wells.

\mysection{Conclusion}
We analyzed the possible mechanisms of the quantum-Hall stripe pinning along the crystallographic
direction [110]---the native effective mass anisotropy of 2D electrons. It was shown that
for a symmetric quantum well this anisotropy was a linear function of applied
per\-pen\-di\-cu\-lar-to-plane electric field. If it is the anisotropy that defines the
direction of the stripes, the critical in-plane magnetic field, when the resistance goes
isotropic, will strongly depend on the bias. The proper experiment will answer the question
whether the native anisotropy of the effective mass is the mechanism of the pinning.

\vspace{0.2cm}
The work was supported by RFBR (99-02-17592 and 01-02-06476), Federal Programs FTNS and PAS.
Besides, E.E.T. was supported in part by the Young Scientists Support Program of RAS
(Grant \#43).

\mysection{References} 
\newlength{\mydescr}
\settowidth{\mydescr}{[1]}
\begin{list}{}%
{\setlength{\labelwidth}{\mydescr}%
\setlength{\leftmargin}{\parindent}%
\setlength{\topsep}{0pt}%
\setlength{\itemsep}{0pt}}
\item [\textrm{[1]}] M.P.~Lilly, et al., Phys.\ Rev.\ Lett. {\bf 82}, 394 (1999);
R.R.~Du, et al., Solid State Commun. {\bf 109}, 389 (1999).
\item [\textrm{[2]}] M.M.~Fogler, et al., Phys.\ Rev.\ B {\bf 54}, 1853 (1996);
M.M.~Fogler and A.A.~Koulakov, ibid {\bf 55}, 9326 (1997);
R.~Moessner and J.T.~Chalker, ibid {\bf 54}, 5006 (1996).
\item [\textrm{[3]}] G.R.~Aizin and V.A.~Volkov, Sov.\ Phys.\ JETP {\bf 60}, 844 (1984);
ibid {\bf 65}, 188 (1987);
A.H.~MacDonald and M.P.A.~Fisher, Phys.\ Rev.\ B {\bf 61}, 5724 (2000);
F.~von~Oppen, et al., Phys.\ Rev.\ Lett. {\bf 84}, 2937 (2000).
\item [\textrm{[4]}] W.~Pan, et al., Phys. Rev. Lett. {\bf 83}, 820 (1999);
M.P.~Lilly, et al., ibid {\bf 83}, 824 (1999).
\item [\textrm{[5]}] T.~Jungwirth, et al., Phys. Rev. B {\bf 60}, 15574 (1999);
T.D.~Stanescu, et al.,  Phys. Rev. Lett. {\bf 84}, 1288 (2000).
\item [\textrm{[6]}] H.~Kroemer, cond-mat/9901016.
\item [\textrm{[7]}] E.E.~Takhtamirov and V.A.~Volkov, JETP Lett. {\bf 71}, 422 (2000),
cond-mat/0006226.
\item [\textrm{[8]}] W.~Pan, et al., Phys. Rev. Lett. {\bf 85}, 3257 (2000).
\item [\textrm{[9]}] E.E.~Takhtamirov and V.A.~Volkov, JETP {\bf 89}, 1000 (1999).
\item [\textrm{[10]}] K.B.~Cooper, et al., cond-mat/0104243.

\end{list} 

\end{document}